\documentclass[submission, Proceedings]{SciPost}

\hypersetup{
    colorlinks,
    linkcolor={red!50!black},
    citecolor={blue!50!black},
    urlcolor={blue!80!black}
}

\usepackage[bitstream-charter]{mathdesign}
\DeclareSymbolFont{usualmathcal}{OMS}{cmsy}{m}{n}
\DeclareSymbolFontAlphabet{\mathcal}{usualmathcal}

\begin{document}

\begin{center}{\Large \textbf{
TAIGA - an advanced hybrid detector complex for astroparticle physics and high energy gamma-ray astronomy\\
}}\end{center}

\begin{center}
N. Budnev\textsuperscript{1,$\star$},
I. Astapov\textsuperscript{2},
P. Bezyazeekov\textsuperscript{1},
E. Bonvech\textsuperscript{3},
A. Borodin\textsuperscript{4},
A. Bulan\textsuperscript{3},
D. Chernov\textsuperscript{3},
A. Chiavassa\textsuperscript{5},
A. Dyachok\textsuperscript{1},
A. Gafarov\textsuperscript{1},
A. Garmash\textsuperscript{6,7},
V. Grebenyuk\textsuperscript{4,8},
E. Gress\textsuperscript{1},
O. Gress\textsuperscript{1},
T. Gress\textsuperscript{1},
A. Grinyuk\textsuperscript{4},
O. Grishin\textsuperscript{1},
A. D. Ivanova\textsuperscript{1},
A. L. Ivanova\textsuperscript{1,7},
N. Kalmykov\textsuperscript{3},
V. Kindin\textsuperscript{2},
S. Kiryuhin\textsuperscript{1},
R. Kokoulin\textsuperscript{2},
K. Kompaniets\textsuperscript{2},
E. Korosteleva\textsuperscript{3},
V. Kozhin\textsuperscript{3},
E. Kravchenko\textsuperscript{6,7},
A. Kryukov\textsuperscript{3},
L. Kuzmichev\textsuperscript{1,3},
A. Lagutin\textsuperscript{9},
M. Lavrova\textsuperscript{4},
Y. Lemeshev\textsuperscript{1},
B. Lubsandorzhiev\textsuperscript{10},
N. Lubsandorzhiev\textsuperscript{3},
A. Lukanov\textsuperscript{10},
D. Lukyantsev\textsuperscript{1},
S. Malakhov\textsuperscript{1},
R. Mirgazov\textsuperscript{1},
R. Monkhoev\textsuperscript{1},
E. Osipova\textsuperscript{3},
A. Pakhorukov\textsuperscript{1},
L. Pankov\textsuperscript{1},
 A. Pan\textsuperscript{4},
   A. Panov\textsuperscript{3},
A. Petrukhin\textsuperscript{2},
I. Poddubnyi\textsuperscript{1},
D. Podgrudkov\textsuperscript{3},
V. Ponomareva\textsuperscript{1},
E. Popova\textsuperscript{3},
E. Postnikov\textsuperscript{3},
V. Prosin\textsuperscript{3},
V. Ptuskin\textsuperscript{11},
A. Pushnin\textsuperscript{1},
R. Raikin\textsuperscript{9},
A. Razumov\textsuperscript{3},
G. Rubtsov\textsuperscript{10},
E. Ryabov\textsuperscript{1},
V. Samoliga\textsuperscript{1},
A. Satyshev\textsuperscript{4},
A. Silaev\textsuperscript{3},
A. Silaev (junior)\textsuperscript{3},
A. Sidorenkov\textsuperscript{10},
A. Skurikhin\textsuperscript{3},
A. Sokolov\textsuperscript{7},
L. Sveshnikova\textsuperscript{3},
V. Tabolenko\textsuperscript{1},
L. Tkachev\textsuperscript{4,8},
A. Tanaev\textsuperscript{1},
M. Ternovoy\textsuperscript{1},
R. Togoo\textsuperscript{12},
N. Ushakov\textsuperscript{10},
A. Vaidyanathan\textsuperscript{7},
P. Volchugov\textsuperscript{3},
N. Volkov\textsuperscript{9},
D. Voronin\textsuperscript{10},
A. Zagorodnikov\textsuperscript{1},
D. Zhurov\textsuperscript{1,13} and
I. Yashin\textsuperscript{2}
\end{center}

\begin{center}
{\bf 1} Irkutsk State University, Karl Marx str. 1, Irkutsk, 664003 Russia
\\
{\bf 2} National Research Nuclear University MEPhI (Moscow Engineering Physics Institute), Kashirskoe hwy 31, Moscow, 115409 Russia
\\
{\bf 3} Skobeltsyn Institute of Nuclear Physics MSU, Leninskie gory 1 (2), GSP-1, Moscow, 119991 Russia
\\
{\bf 4} Joint Institute for Nuclear Research, Joliot-Curie str. 6, Dubna, 141980 Russia
\\
{\bf 5} Dipartimento di Fisica Generale Universiteta di Torino and INFN, Via P. Giuria 1, Turin, 10125 Italy
\\
{\bf 6} Budker Institute of Nuclear Physics SB RAS, Ac. Lavrentiev Avenue 11, Novosibirsk, 630090 Russia
\\
{\bf 7} Novosibirsk State University, Pirogova str. 2, Novosibirsk, 630090 Russia
\\
{\bf 8} Dubna State University, Universitetskaya str. 19, Dubna, 141982 Russia
\\
{\bf 9} Altai State University, pr. Lenina 61, Barnaul, 656049 Russia
\\
{\bf 10} Institute for Nuclear Research RAS, prospekt 60-letiya Oktyabrya 7a, Moscow, 117312 Russia
\\
{\bf 11} IZMIRAN, Kaluzhskoe hwy 4, Troitsk, Moscow, 4108840 Russia
\\
{\bf 12} Institute of Physics and Technology Mongolian Academy of Sciences, Enkhtaivan av. 54B, Ulaanbaatar, 210651 Mongolia
\\
{\bf 13} Irkutsk National Research Technical University, Lermontov str. 83, Irkutsk, 664074 Russia
\\

* nbudnev@api.isu.ru
\end{center}

\begin{center}
\today
\end{center}


\definecolor{palegray}{gray}{0.95}
\begin{center}
\colorbox{palegray}{
  \begin{tabular}{rr}
  \begin{minipage}{0.1\textwidth}
    \includegraphics[width=30mm]{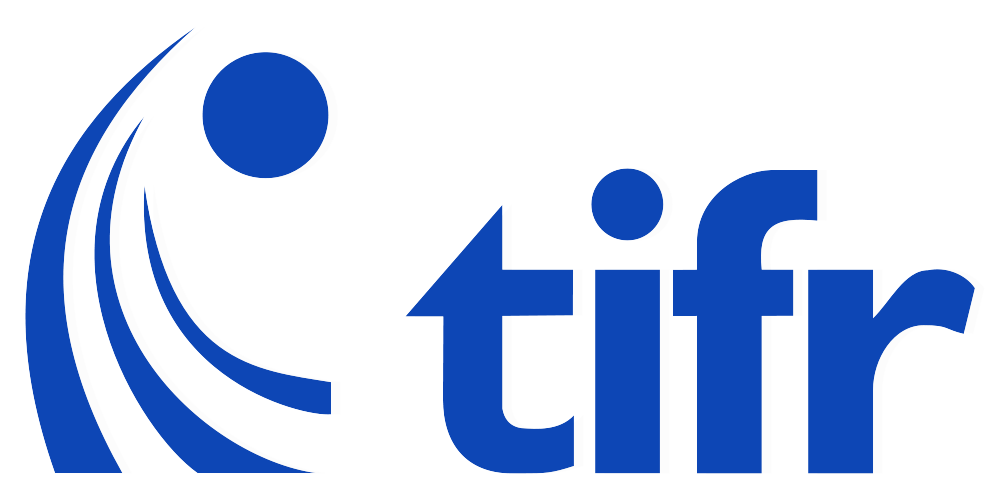}
  \end{minipage}
  &
  \begin{minipage}{0.85\textwidth}
    \begin{center}
    {\it 21st International Symposium on Very High Energy Cosmic Ray Interactions (ISVHE- CRI 2022)}\\
    {\it Online, 23-27 May 2022} \\
    \doi{10.21468/SciPostPhysProc.}\\
    \end{center}
  \end{minipage}
\end{tabular}
}
\end{center}

\section*{Abstract}
{\bf
The physical motivations, present status, main results in study of cosmic rays and in the field of gamma-ray astronomy as well future plans of the TAIGA-1 (Tunka Advanced Instrument for cosmic ray physics and Gamma Astronomy) project are presented. The TAIGA observatory addresses ground-based gamma-ray astronomy and astroparticle physics at energies from a few TeV to several PeV, as well as cosmic ray physics from 100 TeV to several EeV. The pilot TAIGA-1 complex is located in the Tunka valley, ~50 km west from the southern tip of the lake Baikal.
}

\vspace{10pt}
\noindent\rule{\textwidth}{1pt}
\tableofcontents\thispagestyle{fancy}
\noindent\rule{\textwidth}{1pt}
\vspace{10pt}

\section{Introduction}
\label{sec:intro}
The experimental data necessary to understand the nature of the sources of high-energy cosmic rays are obtained both by methods of gamma astronomy and neutrino astrophysics, studying the flux of neutral particles generated by cosmic rays in the immediate vicinity of the source, and by the remote methods of cosmic ray physics, studying the energy spectrum and the mass composition of cosmic rays near the Earth.
The TAIGA(Tunka Advanced Instrument for cosmic rays and Gamma-ray Astronomy)  astrophysical facility \cite{1,2} allows the nature of cosmic ray (CR) sources to be investigated within both gamma-ray astronomy and cosmic ray physics methods. A unique feature of the facility is combining arrays with different types of detectors into a unified system to detect all extensive air shower (EAS) components. This will allow one to search for PeVatrons, i.e., galactic objects in which protons are accelerated to energies $\approx10^{14}$-$10^{17}$ eV, to find the energy limits for particle acceleration in supernova remnants and pulsar nebulae, and to carry out a search for correlations with neutrino events recorded by the Ice Cube \cite{3} and Baikal-GVD \cite{4} neutrino observatories.

\section{The Tunka valley experimental facility}
The experiments to detect EASs by their Cherenkov radiation in the Tunka valley 50 km to the west from Lake Baikal (51.49 N, 103.04 E) were begun in 1993. The first experimental array Tunka-4 consisted of four optical detectors \cite{5} on the base of Hybrid QUASAR-370 phototube, designed for the Baikal neutrino telescope NT200 \cite{6}.
Then in 2012 we finished deployment of the Tunka-133 array consisting of 175 Cherenkov detectors \cite{7,8} arranged over an area of 3 km$^2$. In 2022, we completed the deployment and commissioning of the TAIGA-1 hybrid
complex \cite{9,09}. It consists of 120 TAIGA-HiSCORE wide-angle
Cherenkov stations, three 4-m class Imaging Atmospheric Cherenkov Telescopes
(IACT) and 250 m$^2$ of scintillation particle detectors of Tunka-Grande \cite{10} and TAIGA-MUON \cite{11} arrays.
Status and main results in the field of gamma-ray astronomy are presented in the paper. As well a new approach to the Search for Astrophysical Nanosecond Optical Transients with TAIGA-HiSCORE Array are discussed.

\subsection{Cosmic rays energy spectrum and mass composition}
The energy spectrum of cosmic rays based on the data of HiSCORE and Tunka-133  \cite{9,24}
covers  the energy range from 300 TeV to 3 EeV.  The main result is the proof of a complex energy dependence of the CR intensity that was assumed previously. In addition to the knee at an energy of 3 PeV, statistically significant features are observed at an energy of 20 PeV, at this energy the power law index changes from 3.3 to 3.0, but at an energy of 300 PeV the power law index increases again to 3.3.
A comparison of the TAIGA energy spectrum with other experimental data is presented in \cite{024}. 
At energies smaller than 1 PeV there is good agreement both with
the direct measurements at  balloons \cite{25}, satellite \cite{26},
and high - altitude experiments \cite{27}.

 At high energies it is in agreement with the results of giant arrays (the Pierre Auger Observatory (PAO) \cite{30,37} and the Telescope Array (TA) \cite{31,38}). In all experiments, there is a decrease in the power exponent at an energy of 15-25 PeV by 0.2–0.3. Currently, this effect has no astrophysical explanation.
 
 For the Tunka-Grande array the primary particle energy is reconstructed from the EAS particle flux density at a distance of 200 m ($\rho_{200}$) from the core. Correlation of $\rho_{200}$ with the primary energy is determined using the experimental results of Tunka-133  array \cite{10}.
 At this conference energy spectrum based on five years of Tunka-Grande operation is presented \cite{34}. A change in the power law index at an energy of 20 PeV is also observed according to the Tunka Grande installation.
 
 To measure the cosmic rays mass composition with TAIGA Cherenkov arrays we reconstruct a depth  of the EAS maximum Xmax which depends linearly on ln A for a fixed energy, where A is the atomic mass. The  TAIGA energy dependence of the average value of ln A is given in \cite{034}.
 
\section{Gamma-ray astronomy}
In the past decade, impressive success in the studies of high-energy gamma radiation has been made using third-generation imaging air Cherenkov telescopes (IACTs): H.E.S.S., MAGIC, and VERITAS \cite{40,41,42}. These telescopes are aimed at investigating the energy range of 50 GeV to 50 TeV.

The currently most ambitious CTA gamma-ray observatory project \cite{43} plans to use about 40 small-sized telescopes (SSTs) spread out over an area of 4.5 km$^2$ for exploring the energy range above 10 TeV. Over the past three years, a breakthrough in the range of gamma-ray energies above 100 TeV has been made by high-altitude installations: Tibet AS$\gamma$ \cite{44}, HAWC \cite{45} and LHAASO \cite{46,47}. At the moment, TAIGA is the northernmost gamma-rays observatory at the world level. The observatory's observation program includes sources whose observation time is long enough for the northern location of the observatory: Crab Nebula, Dragonfly, Tycho Brahe supernova remnants, CTA-1, G106.3+2.7, sources in the Sygnys Coocon nebula, blazars Mgk501, Mrk421, etc.

\subsection{Detection of high-energy gamma quanta}
A new approach to the studies of high-energy gamma rays is being developed within the TAIGA installation \cite{7,49} with a system of wide-angle arrays for sampling-timing of the Cherenkov light shower front and IACTs \cite{50}. The TAIGA installations provide us with three different approaches for the detection of high-energy gamma quanta:
\subsubsection{Stand-alone mode of an IACT operation for E $<$ 60 TeV}
To distinguish events from gamma-rays against the background of events from cosmic rays, according to the data of one telescope, differences in the angular distribution of the Cherenkov radiation of the EAS are used, leading to a difference in the shape of the images in the telescope camera. The shape of the image is characterized by the Hillas parameters (width, length, dist, con etc.) which are restored by the values of signal amplitudes in camera pixels. Width and length parameters are standard deviations along the minor and major axis, dist parameter is a distance from the weighted centroid of the image to the source, con is a concentration parameter of the image. A detailed description of these and other Hillas parameters is given in  \cite{51,52}. Using the system of cuts to the Hillas parameters \cite{51} or the widely used Random Forest Classifier \cite{53,54}, it is possible to suppress the number of events from hadrons to a level sufficient to isolate events from gamma quanta. The effective area of registration of events from gamma quanta, after applying the selection criteria, is $5\cdot10^4$ m$^2$ at energies above 10 TeV. The next important parameter we used for background suppression is the angle $\Theta$ between  the arrival direction of the shower and the direction to the source. The effective algorithm for the reconstruction of the arrival direction of TeV gamma rays using a single telescope was proposed in \cite{55} and later was developed as ‘disp’ method \cite{56,57}.

The energy of EAS can be determined by the total number of photoelectrons detected by IACT and the distance from the EAS core, calculated by Hillas parameters described in \cite{51,52,58}. Hillas parameter dist can be used to measure such a distance for the EAS from gamma rays. It was shown that for a single IACT it is possible to achieve an accuracy of energy reconstruction of around 30\% in the TeV energy range. Such accuracy is sufficient to reconstruct the spectra of gamma rays.

Observations of the source were carried out in wobbling mode, when the position of the source on the camera was shifted by 1.2 degrees relative to the center of the camera. The Hillas parameters (called ‘On’) are calculated relative to this position, and the parameters of background events (called ‘Off’) are calculated relative to the anti-source located at a distance of 1.2 degrees from the center, but in the opposite direction.

\subsubsection{Stereoscopic approach for large distances between the IACTs for E \textgreater 10 TeV}
When the energy of EAS is higher than 10 TeV,  EAS may be detected by two or more telescopes. It becomes possible to use a stereoscopic approach for selecting events from gamma-ray. This approach has shown its effectiveness in modern gamma-ray observatories (MAGIC, H.E.S.S., VERITAS) with a distance between telescopes up to 100 m. The stereoscopic approach is also planned to use in CTA. In this experiment small telescopes will be located at a distance of 250 m. An example of the image of the EAS detected by our three telescopes is shown in Fig. 1.
When detecting an event with two telescopes, the reconstruction accuracy of the EAS core position is better than 20 m. For 68\% of events from gamma-rays, the reconstructed arrival direction of EAS differs from the direction to the source by an angle which is less than 0.25 degrees. The accuracy of energy recovery is about 10\%. The effective area of registration of events from gamma quanta, after applying the selection criteria, is $5\cdot10^5$ m$^2$ at energy above 40 TeV, which is 2 times larger than that of three telescopes in the stand-alone mode.
\begin{figure}[t]
	\centering
	\includegraphics[width=1\textwidth]{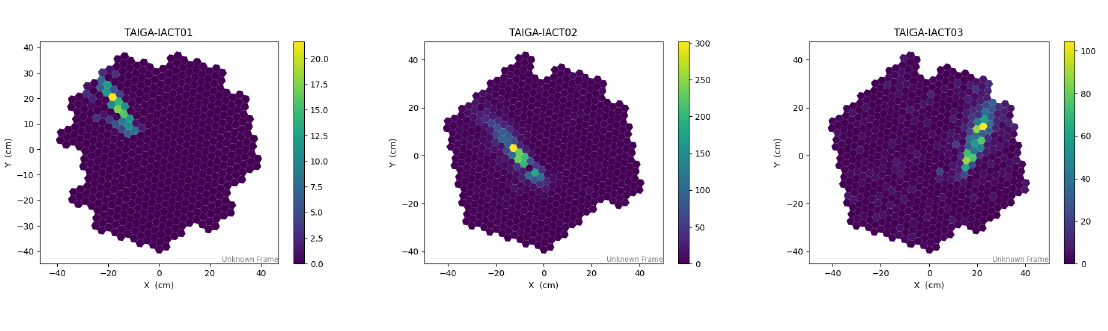}
	\caption{An example of an event recorded by three telescopes.}
	\label{ref}
\end{figure}

\subsubsection{Hybrid approach - joint operation of TAIGA-HiSCORE and IACTs for E >40 TeV}
At energies above 40 TeV, a new "hybrid" approach for gamma-ray detection becomes possible - the detection of EAS by both IACTs and the TAIGA-HiSCORE installation. The main advantage of the operation of the IACTs in the network of wide-angle Cherenkov stations is a more efficient selection of events from gamma-rays from the background of charged cosmic rays. EAS image parameters (Hillas parameters) are supplemented by the parameters of the EAS (core position, arrival direction, energy), which are well reconstructed by the TAIGA-HiSCORE wide-angle installation. With the joint work of the first telescope and one-quarter of the TAIGA-HiSCORE installation, four events from gamma quanta with energies above 100 TeV were identified during 100 hours of observation of the Crab Nebula. With the entire installation of TAIGA-HiSCORE and three telescopes, 20-25 events with an energy above 100 TeV are expected for 100 hours of observation.
\subsection{Gamma-rays from Crab Nebular}
The gamma-ray source in the Crab Nebula was observed by the first atmospheric Cherenkov telescope for 150 hours during two seasons (2019-2020 and 2020-2021), 567 events ($\Theta^2$ <0.05) from gamma-rays in the energy range of 5-100 TeV were selected (Fig 2, left). The significance level of this number of events is 11 $\sigma$.
Figure 2, right shows the energy spectrum of gamma-rays obtained using the energy reconstruction technique briefly described above. The obtained spectrum of particles coincides quite well with the world data in the range from 5 to 100 TeV.
During the first 36 hours of the operation of the 1st and 2nd IACTs in stereo mode, a signal at the significance level of 5 $\sigma$ was obtained, and the energy spectrum of these events was reconstructed. Obtained spectrum is in good agreement with other experiments.

\begin{figure}[t]
	\centering
	\includegraphics[width=1\textwidth]{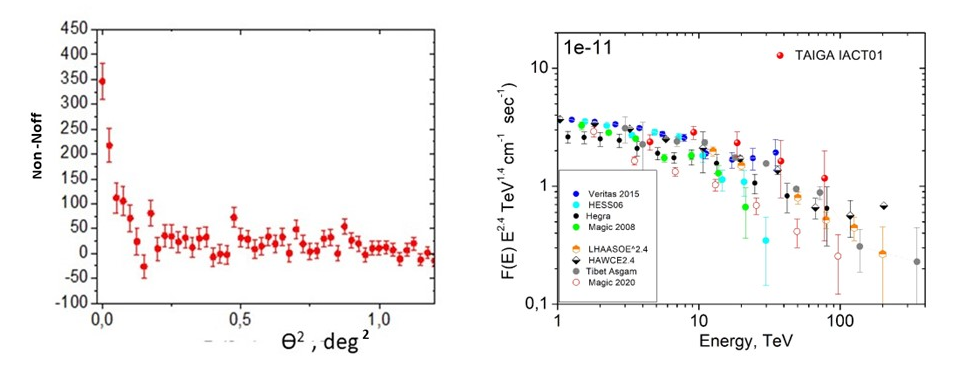}
	\caption{Background subtracted $\Theta^2$-distribution (left). Energy spectrum of gamma quanta from the Crab nebula according to TAIGA-IACT (red circle) in comparison with the results of other experiments (right)}
	\label{ref}
\end{figure}

\section{Search for astrophysical nanosecond optical transients with TAIGA-HiSCORE array}
The stations of the TAIGA-HiSCORE installation are adapted for detecting optical pulses of Cherenkov light with a duration from units of ns to several tens of ns. It was shown \cite{62} that the events of remote point optical transients of any nature can be distinguished from the events of the EAS (gamma quanta or cosmic rays) by the structure of the events of the HiSCORE installation: the events of the EAS have the form of a compact set of installation stations, but the events of remote point sources illuminate the entire installation uniformly. Due to this, it is possible to organize event filtering in the HiSCORE data stream to search for possible events of remote optical transients of astrophysical origin. Compared to conventional astronomical and radio astronomy instruments with a very narrow aperture, a distinctive feature of the HiSCORE installation, which is extremely useful for searching for rare optical transients, is a wide field of view, which is about $60^o$ with a sufficiently high angular resolution. Currently, the HiSCORE data processing technique has already been developed and tested to search for optical nanosecond transients and one complete winter observation season (2018-2019) has been processed.

\section{Conclusion}
The TAIGA-1 astrophysical complex will make it possible to advance in solving a number of topical problems of gamma-ray astronomy and cosmic ray physics. It is planned to reconstruct the gamma-ray energy spectrum from galactic sources such as the Crab Nebula, Dragonfly, J2227+610 (G106.3+2.7), J2031 +415 (Cygnus Cocoon) and the Tycho Brahe supernova. It is important for understanding the origin of cosmic-rays, to search for high-energy gamma-rays associated with high-energy neutrinos detected by the IceCube and Baikal-GVD neutrino telescopes and new sources of high-energy gamma-rays. About 60 gamma-quanta with energies above 100 TeV are expected for the Crab Nebula gamma-ray source in 300 hours of observation.
The study of the mass composition of cosmic rays will continue according to the Tunka-133, TAIGA-HiSCORE and Tunka-Grande installations at energies above $10^{15}$ eV.
An attempt will be made to study the mass composition at energies below $10^{15}$ eV, possibly, based on the joint TAIGA-HiSCORE and TAIGA-IACT data. In the range $10^{16}$–$10^{18}$ eV we hope to advance in studying the mass composition through an increase in the area of the muon detectors.
The next plans for the development of the TAIGA-1 installation include the deployment in 2023-2024 of 2 more IACTs and 200 m$^2$ of particle detectors.
In the future, it is planned to use Small Imaging Telescopes (SITs) with cameras with a FOV of 25-30$^\circ$ and an effective detection area of 1 m$^2$ to study the energy range above 80 TeV. With the operation of such telescopes, the percentage of joint events with the TAIGA-HISCORE installation will increase by almost 10 times and for joint events, the high efficiency of separating events from gamma quanta will remain.
The further development of the TAIGA project is mainly due to the expansion of the TAIGA-HISCORE installation. With a ten-fold increase in area (with the creation of the TAIGA-10 setup), the number of events from the Crab Nebula for 100 hours of observation will reach 300 at energies above 100 TeV, with the significance of $\approx$ 5 $\sigma$. Additional suppression of hadron background using wide angle SITs will increase this value to $\approx$ 10 $\sigma$.

\section*{Acknowledgements}
The work was performed at the UNU “Astrophysical Complex of MSU-ISU” (agreement EB 075-15-2021-675). The work is supported by RFBR (grant 19-32-60003), the RSF(grants 19-72-20067(Section 4)), the Russian Federation Ministry of Science and High Education (projects FZZE-2020-0017, FZZE-2020-0024, FZZE-2022-0001 and FSUS-2020-0039).




\nolinenumbers

\end{document}